\documentclass{article}

\usepackage{PRIMEarxiv}

\usepackage{abstract}
\usepackage[utf8]{inputenc}
\usepackage[T1]{fontenc}
\usepackage{hyperref}
\usepackage{url}
\usepackage{booktabs}
\usepackage{array}
\usepackage{tabularx}
\usepackage{xcolor}
\usepackage[normalem]{ulem}
\usepackage{amsfonts}
\usepackage{amsmath}
\usepackage{amsthm}
\usepackage{nicefrac}
\usepackage{microtype}
\usepackage{fancyhdr}
\usepackage{graphicx}
\usepackage{placeins}
\usepackage{tikz}
\usetikzlibrary{arrows.meta,positioning}
\graphicspath{{media/}}

\definecolor{archsoftblue}{RGB}{235,244,250}
\definecolor{archsofttan}{RGB}{252,246,232}
\definecolor{archsoftgreen}{RGB}{237,248,239}
\definecolor{archsoftgray}{RGB}{246,246,246}
\definecolor{archsoftyellow}{RGB}{255,249,229}

\tikzset{
  archbox/.style={
    draw=black!58,
    rounded corners=1.5pt,
    line width=0.45pt,
    align=left,
    inner sep=5pt,
    font=\sffamily\fontsize{7.2}{8.1}\selectfont
  },
  archcorebox/.style={archbox, line width=0.75pt},
  archarrow/.style={-{Latex[length=2.4mm,width=1.8mm]}, line width=0.55pt, black!75, shorten >=2.5pt, shorten <=2.5pt},
  archline/.style={line width=0.55pt, black!75, shorten <=2.5pt},
  archbranch/.style={-{Latex[length=2.4mm,width=1.8mm]}, line width=0.55pt, black!75, shorten >=2.5pt},
  archoptional/.style={archarrow, dashed},
  archarrowlabel/.style={font=\sffamily\fontsize{6.4}{6.8}\selectfont, fill=white, inner sep=1.1pt, text=black!70}
}

\hypersetup{
  colorlinks=true,
  linkcolor=black,
  citecolor=black,
  urlcolor=blue,
  pdftitle={Visual-to-Code Authoring, Tensor-Network Debugging, and Quantum-Circuit Inspection Tools in Python},
  pdfauthor={Alejandro Mata Ali},
  pdfsubject={Tensor-network and quantum-circuit workflow tools},
  pdfkeywords={tensor networks, quantum circuits, visualization, Python, scientific software}
}
\newcommand{\toolname}[1]{\textsc{#1}}
\newcommand{\code}[1]{\texttt{#1}}
\newcolumntype{L}[1]{>{\raggedright\arraybackslash}p{#1}}
\newcolumntype{M}[1]{>{\raggedright\arraybackslash}m{#1}}
\newcommand{\pkgsha}[1]{\nolinkurl{#1}}

\newcommand{\revdel}[1]{\textcolor{red}{\sout{#1}}}

\title{Visual-to-Code Authoring, Tensor-Network Debugging, and Quantum-Circuit Inspection Tools in Python}

\author{
  Alejandro Mata Ali \\
  Instituto Tecnol\'ogico de Castilla y Le\'on, Burgos, Spain\\
  \texttt{alejandro.mata.ali@gmail.com} \\
  ORCID: \href{https://orcid.org/0009-0006-7289-8827}{0009-0006-7289-8827}
}

\begin{document}
\maketitle

\begin{abstract}
Tensor networks and quantum circuits are structural objects whose meaning depends on connectivity, indices, contraction order, gate placement, measurements, and related design choices.
They are often easier to reason about visually than as code, yet in Python they are frequently constructed, transformed, and checked through backend-specific objects or compact symbolic expressions.
This can make structural mistakes hard to notice during development, debugging, and communication.
This paper presents three complementary packages: \toolname{Tensor-Network-Visualization} for visual debugging and structural inspection of supported tensor-network and traced \code{einsum} workflows; \toolname{Tensor-Network-Editor} for visual-to-code authoring, backend code generation, JSON preservation, export, and design-level analysis; and \toolname{Quantum Circuit Drawer} for clear circuit rendering, inspection, and complementary comparison of circuits or documented result distributions.
The packages form a visual authoring and inspection layer around existing tensor-network libraries, array-based scientific Python workflows, and quantum SDKs.
They are not simulators: they do not implement new contraction algorithms, execute quantum circuits, or guarantee full semantic equivalence across arbitrary backends.
Their contribution is to make structural artifacts visible, editable, inspectable, comparable, exportable, and reproducible within those ecosystems.
\end{abstract}

\section{Summary}

Tensor networks and quantum circuits are usually easier to reason about as diagrams than as code alone.
In practice, however, researchers often construct them through backend-specific Python objects, compact \code{einsum} expressions, or framework-dependent circuit classes.
That gap between the visual mathematical object and the executable program can make structural mistakes hard to notice: a network may be connected incorrectly, a contraction may not follow the intended scheme, an intermediate tensor may contain unexpected patterns, or two circuits may look similar in code while differing visibly in layout, decomposition, or measurement structure.

The three packages discussed here address this visual gap from complementary directions.
\toolname{Tensor-Network-Visualization} \cite{dokostayos_tnv_2026} focuses on visual debugging and structural inspection of supported tensor-network and traced-\code{einsum} inputs.
\toolname{Tensor-Network-Editor} \cite{dokostayos_tne_2026} focuses on visual-to-code authoring for tensor-network designs, especially when custom or non-standard structures are easier to create visually and then export as backend code.
\toolname{Quantum Circuit Drawer} \cite{dokostayos_qcd_2026} focuses on clear circuit rendering and inspection, with comparison and result-distribution views available for documented inputs.

Together, they sit alongside existing tensor-network libraries and quantum SDKs rather than replacing their computational roles.

\section{Statement of Need}

In Python workflows, tensor networks and quantum circuits are frequently created, transformed, and checked through backend-specific objects or compact symbolic expressions, even though their meaning depends on connectivity, indices, contraction order, gate placement, measurements, decomposition choices, and, in some workflows, hardware topology.
This creates three recurring practical problems: complex tensor-network implementations are cumbersome and error-prone to write directly as code; tensor networks that have already been programmed can be difficult to debug or analyze structurally; and quantum circuits can be awkward to inspect clearly once they become large, transformed, or distributed across framework-specific representations.

For tensor-network workflows, this creates a recurring debugging problem.
A user may want to know whether a network has been connected as intended, whether a traced \code{einsum} expression matches the expected graph, whether a contraction plan follows the desired scheme, or whether an intermediate or resulting tensor contains visible patterns such as concentration, sparsity, or other anomalies.
Existing computational libraries provide powerful modeling and contraction functionality, but visual inspection is often secondary to their numerical role and tied to backend-specific representations.
As a result, structural verification may require ad hoc plotting, custom notebook code, or repeated translation of the same object into several forms before it becomes visually understandable.
This is especially relevant when tensor networks are embedded in machine-learning or numerical workflows: TensorKrowch integrates tensor networks with PyTorch, while many prototype and interoperability paths use PyTorch or NumPy tensors and \code{einsum} expressions rather than a dedicated tensor-network class.

The authoring problem is different but closely related.
Complex or non-standard tensor networks are frequently visual objects first and code artifacts second.
Directly implementing them in Python can therefore be slow and error-prone, especially when the intended structure does not fit a standard family such as MPS, MPO, or PEPS, or when the workflow relies on dense \code{einsum} expressions that are compact but not visually transparent.
What is missing is an authoring layer that lets a user construct the network visually and generate backend code from that structure, reducing manual boilerplate and lowering the chance of structural implementation errors before secondary tasks such as export or preservation are considered.

Quantum-circuit workflows exhibit analogous friction.
Researchers often need a consistent way to inspect circuits visually, navigate large circuits in pages or other managed views, examine decompositions, compare two related implementations, view metadata associated with a gate placement, or inspect result distributions side by side.
When supported by the underlying package, topology-aware layouts can also matter for understanding how a circuit relates to a chip or coupling graph.

This motivates three complementary tools for debugging tensor-network
structure and contents, authoring tensor-network code from visual designs,
and inspecting quantum circuits, with export, comparison, and preservation
available where documented.

\section{Related Software and State of the Field}

The tools in this paper sit near, but not inside, mature computational libraries.
On the tensor-network side, Quimb, TensorNetwork, TensorKrowch, and TeNPy provide tensor-network capabilities for simulation, contraction, physics workflows, or machine-learning integration \cite{gray2018quimb,roberts2019tensornetwork,pareja2024tensorkrowch,hauschild2018tenpy}.
PyTorch and NumPy are not tensor-network libraries in the same sense, but they are important in practice because tensor-network workflows are often embedded in machine-learning pipelines, array programs, and traced \code{einsum} expressions \cite{paszke2019pytorch,harris2020array}.
On the quantum-circuit side, Qiskit, Cirq, PennyLane, CUDA-Q, and MyQLM provide circuit construction, execution, compilation, and ecosystem tooling, differentiation, quantum-classical programming, emulation, or ecosystem-specific abstractions \cite{javadiabhari2024qiskit,cirq_developers_2025,bergholm2018pennylane,nvidia_cuda_q_2026,bull_myqlm_2026}.
Matplotlib \cite{hunter2007matplotlib}, NetworkX \cite{hagberg2008networkx}, and \code{opt\_einsum} \cite{smith2018opteinsum} provide foundational scientific-Python components used directly or conceptually in these workflows.

LaTeX, TikZ, Mermaid, DOT, SVG, PNG, and PDF outputs are treated here as insertion and communication formats rather than as direct computational competitors.
For many papers, notebooks, reports, and teaching materials, the appropriate output is precisely a static figure or a textual snippet that can be inserted into the surrounding document system.
The distinction is that those formats preserve a visual artifact, while the packages discussed here recover structure from supported computational representations for inspection, editing, and export before figures are inserted.

Tables~\ref{tab:tn-framework-scope} and~\ref{tab:qc-framework-scope} separate the two ecosystems instead of mixing tensor-network libraries, quantum SDKs, and figure formats in a single comparison.
The tables are not rankings.
They highlight the kinds of functionality that matter for \toolname{Tensor-Network-Visualization}, \toolname{Tensor-Network-Editor}, and \toolname{Quantum Circuit Drawer}.
The qualitative labels in Tables~\ref{tab:tn-framework-scope} and~\ref{tab:qc-framework-scope} are used operationally rather than evaluatively. ``Primary'' means that the capability is a documented first-class workflow or central package role. ``Partial'' means that the capability is documented or practically possible but is backend-dependent, limited in scope, or usually mediated by user code rather than presented as the package's main role. ``User code'' means that the task can generally be achieved by scripting around the library but is not exposed as the package's own documented workflow. ``Supported subset'' means an explicitly documented subset of inputs or operations. ``Adapter import'' means that the package can import or normalize supported external objects into its own internal representation, without implying full semantic preservation. ``No'' means that the feature is not a documented role of the package in the context considered here.

The distinction is not that existing libraries lack valuable visualization capabilities; many of them do.
Rather, visualization is usually secondary to their computational role and often tied to a specific backend representation.
The packages presented here invert that priority for the supported workflows: structural visibility and manipulation are primary, while computation, simulation, differentiation, execution, and optimization remain with existing numerical and quantum-computing libraries.

\begin{table}[!htbp]
\centering
\caption{Tensor-network ecosystem context for visualization and visual-to-code authoring. Entries are qualitative and refer to typical documented use, not to an exhaustive audit of every extension or notebook workflow. TN, TK, TNV, and TNE denote TensorNetwork, TensorKrowch, \toolname{Tensor-Network-Visualization}, and \toolname{Tensor-Network-Editor}, respectively. PyTorch and NumPy are included because supported workflows may use tensors, arrays, or \texttt{einsum} traces directly, especially in machine-learning and interoperability contexts.}
\begingroup
\scriptsize
\setlength{\tabcolsep}{1pt}
\renewcommand{\arraystretch}{1.15}
\begin{tabular}{@{}L{0.19\linewidth} L{0.085\linewidth} L{0.085\linewidth} L{0.085\linewidth} L{0.085\linewidth} L{0.085\linewidth} L{0.085\linewidth} L{0.12\linewidth} L{0.12\linewidth}@{}}
\toprule
Feature relevant to tensor-network editing or visualization & Quimb & TN & TK & TeNPy & PyTorch & NumPy & TNV & TNE \\
\midrule
Native tensor-network object model & Yes & Yes & Yes, PyTorch-based & Yes & No, tensors only & No, arrays only & Normalized snapshots & Visual design model \\
Contraction or simulation context & Primary & Primary & Training and contraction & Primary & Via tensor operations & Via arrays and \code{einsum} & Contraction inspection only; not a contraction engine & Design/planning support; not a simulator \\
Machine-learning or autodiff integration & Partial & Partial & Primary & Limited & Primary & No autodiff & Via supported inputs & Via generated code \\
Raw tensor or \code{einsum} workflows & Partial & Partial & Via PyTorch tensors & Limited & Primary substrate & Primary substrate & Supported traces & Export/code target \\
Built-in structural visualization & Partial & Partial & Limited & Partial & No & No & Primary & Visual editor \\
Tensor-value inspection in the same workflow & Partial & Partial & Partial & Partial & User code & User code & Primary for supported inputs & Not primary \\
Visual authoring or backend code generation & Not primary & Not primary & Not primary & Not primary & No & No & Translation aids & Primary \\
Exportable figures or document snippets & Partial / user code & Partial / user code & Partial / user code & Partial / user code & User code & User code & Supported & Primary support \\
\bottomrule
\end{tabular}
\endgroup
\label{tab:tn-framework-scope}
\end{table}

\begin{table}[!htbp]
\centering
\caption{Quantum-computing ecosystem context for circuit drawing and inspection. Entries are qualitative and focus on capabilities that affect visualization, inspection, comparison, and publication-ready artifacts rather than on numerical performance or hardware access. QCD denotes \toolname{Quantum Circuit Drawer}.\protect\footnotemark[1]}
\begingroup
\scriptsize
\setlength{\tabcolsep}{1pt}
\renewcommand{\arraystretch}{1.15}
\begin{tabular}{@{}L{0.23\linewidth} L{0.115\linewidth} L{0.115\linewidth} L{0.115\linewidth} L{0.115\linewidth} L{0.115\linewidth} L{0.13\linewidth}@{}}
\toprule
Feature relevant to circuit drawing or inspection & Qiskit & PennyLane & Cirq & CUDA-Q & MyQLM & QCD \\
\midrule
Circuit construction model & Primary & Primary & Primary & Kernel/program model & Primary & Internal model and adapters \\
Simulation, execution, or hardware-facing workflow & Primary & Primary for hybrid workflows & Primary / ecosystem-specific & Primary, GPU/QPU-oriented & Primary emulation workflow & No \\
Compilation, transpilation, or connector support & Primary & Partial & Partial & Compiler-oriented & Plugins and connectors & Adapter import \\
Hybrid or differentiable workflows & Partial & Primary & Limited & Hybrid quantum-classical & Variational and plugin workflows & No \\
Built-in drawing or rendering & Yes & Yes & Yes & Not primary & Partial & Primary \\
Large-circuit, decomposition, or topology inspection & Partial & Partial & Partial & Limited / backend-specific & Partial & Primary for supported inputs \\
Circuit or result-distribution comparison artifacts & Partial / user code & Partial / user code & Partial / user code & User code & User code & Supported subset \\
Exportable figures or document snippets & Partial / user code & Partial / user code & Partial / user code & User code & User code & Primary support \\
\bottomrule
\end{tabular}
\endgroup
\label{tab:qc-framework-scope}
\end{table}
\footnotetext[1]{For QCD, adapter support is intentionally described as scoped support: strong documented paths are distinguished from best-effort or platform-qualified paths, and adapter import does not imply complete preservation of every framework-specific construct.}
\FloatBarrier

\section{Software Design and Architecture}

The design is adapter-based, but the adapters are a means rather than the central result.
Their purpose is to extract the structural information needed for visual inspection, authoring, export, and comparison from several documented input types.
Backend-specific tensor-network objects, saved design files, OpenQASM inputs, and result mappings therefore enter through adapters or parser paths that expose graph topology, visible indices, tensor labels, circuit layers, gate labels, measurement outcomes, layout hints, or JSON design records where those concepts are available.
The resulting normalized representation is then consumed by rendering, diagnostics, export, comparison, and code-generation layers where implemented.
Generated artifacts include static and interactive figures, selected backend code snippets, JSON designs, TikZ/LaTeX-oriented snippets, DOT or Mermaid graphs, and histogram outputs.
Optional backend dependencies are loaded only when the corresponding adapter is used, which helps keep lightweight inspection workflows separate from heavier computational environments.

In this paper, ``backend-agnostic'' therefore refers to structural normalization for supported visualization and authoring tasks.
It does not imply universal semantic equivalence.
The packages do not claim that every backend-specific construct can be preserved exactly, nor that arbitrary custom gates, tensor metadata, optimizer state, execution configuration, or hardware calibration data can be reconstructed from the normalized form.

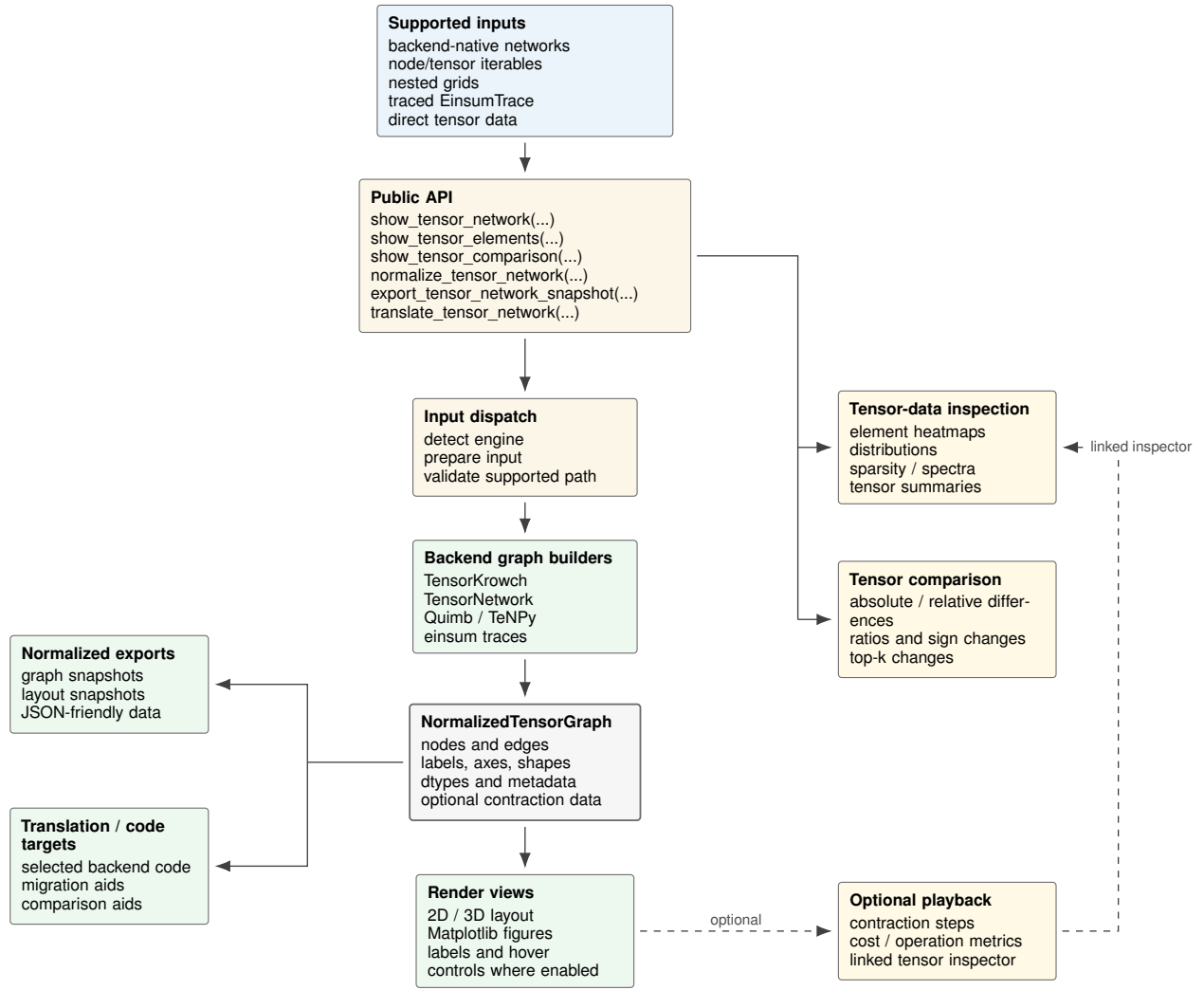
\begin{figure}[!htbp]
  \centering
  \resizebox{\linewidth}{!}{%
  \begin{tikzpicture}[x=1cm,y=1cm]
  \node[archbox, fill=archsoftblue, text width=4.2cm] (f1inputs) at (0,13.2)
    {\textbf{Supported inputs}\\[0.6mm]
    backend-native networks\\
    node/tensor iterables\\
    nested grids\\
    traced EinsumTrace\\
    direct tensor data};

  \node[archbox, fill=archsofttan, text width=4.75cm] (f1api) at (0,10.35)
    {\textbf{Public API}\\[0.6mm]
    show\_tensor\_network(...)\\
    show\_tensor\_elements(...)\\
    show\_tensor\_comparison(...)\\
    normalize\_tensor\_network(...)\\
    export\_tensor\_network\_snapshot(...)\\
    translate\_tensor\_network(...)};

  \node[archbox, fill=archsofttan, text width=3.1cm] (f1dispatch) at (0,7.4)
    {\textbf{Input dispatch}\\[0.6mm]
    detect engine\\
    prepare input\\
    validate supported path};

  \node[archbox, fill=archsoftgreen, text width=3.1cm] (f1builders) at (0,5.1)
    {\textbf{Backend graph builders}\\[0.6mm]
    TensorKrowch\\
    TensorNetwork\\
    Quimb / TeNPy\\
    einsum traces};

  \node[archcorebox, fill=archsoftgray, text width=3.2cm] (f1graph) at (0,2.55)
    {\textbf{NormalizedTensorGraph}\\[0.6mm]
    nodes and edges\\
    labels, axes, shapes\\
    dtypes and metadata\\
    optional contraction data};

  \node[archbox, fill=archsoftgreen, text width=3.0cm] (f1render) at (0,-0.05)
    {\textbf{Render views}\\[0.6mm]
    2D / 3D layout\\
    Matplotlib figures\\
    labels and hover\\
    controls where enabled};

  \node[archbox, fill=archsoftgreen, text width=2.7cm] (f1exports) at (-6.4,3.75)
    {\textbf{Normalized exports}\\[0.6mm]
    graph snapshots\\
    layout snapshots\\
    JSON-friendly data};

  \node[archbox, fill=archsoftgreen, text width=2.7cm] (f1translation) at (-6.4,0.95)
    {\textbf{Translation / code targets}\\[0.6mm]
    selected backend code\\
    migration aids\\
    comparison aids};

  \node[archbox, fill=archsoftyellow, text width=3.0cm] (f1inspect) at (6.5,7.4)
    {\textbf{Tensor-data inspection}\\[0.6mm]
    element heatmaps\\
    distributions\\
    sparsity / spectra\\
    tensor summaries};

  \node[archbox, fill=archsoftyellow, text width=3.0cm] (f1compare) at (6.5,4.75)
    {\textbf{Tensor comparison}\\[0.6mm]
    absolute / relative differences\\
    ratios and sign changes\\
    top-k changes};

  \node[archbox, fill=archsoftyellow, text width=3.0cm] (f1playback) at (6.5,-0.05)
    {\textbf{Optional playback}\\[0.6mm]
    contraction steps\\
    cost / operation metrics\\
    linked tensor inspector};

  \draw[archarrow] (f1inputs.south) -- (f1api.north);
  \draw[archarrow] (f1api.south) -- (f1dispatch.north);
  \draw[archarrow] (f1dispatch.south) -- (f1builders.north);
  \draw[archarrow] (f1builders.south) -- (f1graph.north);
  \draw[archarrow] (f1graph.south) -- (f1render.north);
  \draw[archoptional] (f1render.east) -- node[archarrowlabel, above] {optional} (f1playback.west);
  \draw[archline] (f1graph.west) -- (-3.35,2.55);
  \draw[archbranch] (-3.35,2.55) |- (f1exports.east);
  \draw[archbranch] (-3.35,2.55) |- (f1translation.east);
  \draw[archline] (f1api.east) -- (4.2,10.35) -- (4.2,4.75);
  \draw[archbranch] (4.2,7.4) -- (f1inspect.west);
  \draw[archbranch] (4.2,4.75) -- (f1compare.west);
  \draw[archoptional] (f1playback.east) -- ++(0.95,0) |- node[archarrowlabel, near end, right] {linked inspector} (f1inspect.east);
  \end{tikzpicture}%
  }
  \caption[Architecture of tensor-network-visualization]{Architecture of \texttt{tensor-network-visualization}. Supported backend-native networks, traced einsum workflows, and direct tensor data enter through public inspection, normalization, rendering, export, and translation APIs. Structural inputs are dispatched through backend graph builders into a normalized tensor graph for 2D/3D rendering, optional contraction playback, snapshot export, and selected translation targets, while tensor-data inspection and tensor comparison form parallel public-API branches.}
  \label{fig:architecture-tn-visualization}
\end{figure}

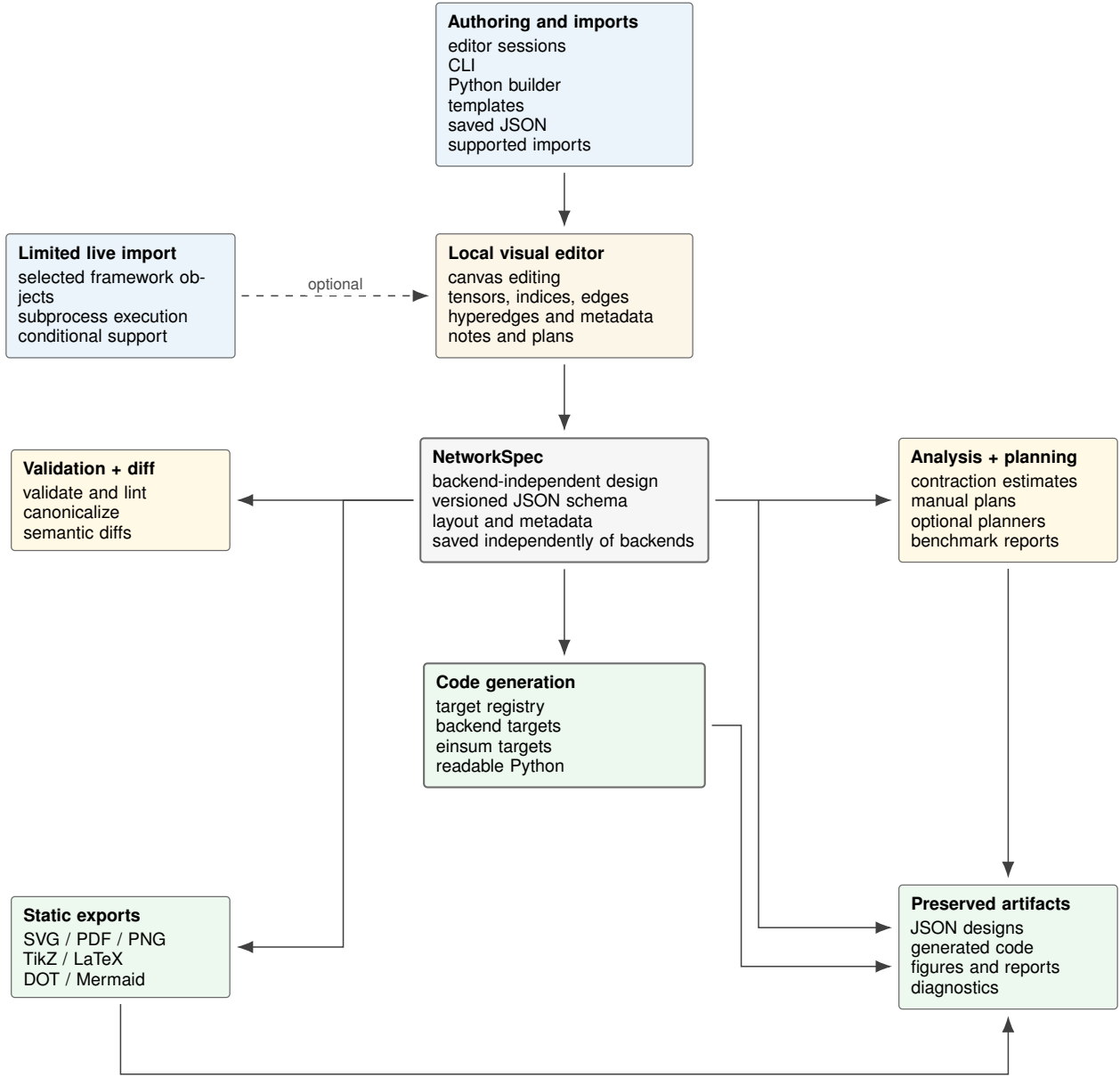
\begin{figure}[!htbp]
  \centering
  \resizebox{\linewidth}{!}{%
  \begin{tikzpicture}[x=1cm,y=1cm]
  \node[archbox, fill=archsoftblue, text width=3.35cm] (f2authoring) at (0,12.7)
    {\textbf{Authoring and imports}\\[0.6mm]
    editor sessions\\
    CLI\\
    Python builder\\
    templates\\
    saved JSON\\
    supported imports};

  \node[archbox, fill=archsoftblue, text width=2.95cm] (f2live) at (-6.4,9.65)
    {\textbf{Limited live import}\\[0.6mm]
    selected framework objects\\
    subprocess execution\\
    conditional support};

  \node[archbox, fill=archsofttan, text width=3.35cm] (f2editor) at (0,9.65)
    {\textbf{Local visual editor}\\[0.6mm]
    canvas editing\\
    tensors, indices, edges\\
    hyperedges and metadata\\
    notes and plans};

  \node[archcorebox, fill=archsoftgray, text width=3.8cm] (f2spec) at (0,6.7)
    {\textbf{NetworkSpec}\\[0.6mm]
    backend-independent design\\
    versioned JSON schema\\
    layout and metadata\\
    saved independently of backends};

  \node[archbox, fill=archsoftyellow, text width=2.8cm] (f2validation) at (-6.4,6.7)
    {\textbf{Validation + diff}\\[0.6mm]
    validate and lint\\
    canonicalize\\
    semantic diffs};

  \node[archbox, fill=archsoftyellow, text width=2.8cm] (f2analysis) at (6.4,6.7)
    {\textbf{Analysis + planning}\\[0.6mm]
    contraction estimates\\
    manual plans\\
    optional planners\\
    benchmark reports};

  \node[archbox, fill=archsoftgreen, line width=0.75pt, text width=3.7cm] (f2codegen) at (0,3.45)
    {\textbf{Code generation}\\[0.6mm]
    target registry\\
    backend targets\\
    einsum targets\\
    readable Python};

  \node[archbox, fill=archsoftgreen, text width=2.8cm] (f2exports) at (-6.4,0.25)
    {\textbf{Static exports}\\[0.6mm]
    SVG / PDF / PNG\\
    TikZ / LaTeX\\
    DOT / Mermaid};

  \node[archbox, fill=archsoftgreen, text width=2.8cm] (f2preservation) at (6.4,0.25)
    {\textbf{Preserved artifacts}\\[0.6mm]
    JSON designs\\
    generated code\\
    figures and reports\\
    diagnostics};

  \draw[archarrow] (f2authoring.south) -- (f2editor.north);
  \draw[archoptional] (f2live.east) -- node[archarrowlabel, above] {optional} (f2editor.west);
  \draw[archarrow] (f2editor.south) -- (f2spec.north);
  \draw[archarrow] (f2spec.south) -- (f2codegen.north);
  \draw[archarrow] (f2spec.west) -- (f2validation.east);
  \draw[archarrow] (f2spec.east) -- (f2analysis.west);
  \draw[archarrow] (f2spec.west) -- ++(-1.1,0) |- (f2exports.east);
  \draw[archbranch] (2.8,6.7) -- (2.8,0.53) -- ([xshift=-0.7cm,yshift=2.8mm]f2preservation.west) -- ([yshift=2.8mm]f2preservation.west);
  \draw[archarrow] (f2codegen.east) -- (2.55,3.45) -- (2.55,-0.03) -- ([xshift=-0.7cm,yshift=-2.8mm]f2preservation.west) -- ([yshift=-2.8mm]f2preservation.west);
  \draw[archarrow] (f2analysis.south) -- (f2preservation.north);
  \draw[archarrow] (f2exports.south) -- ++(0,-1.1) -| (f2preservation.south);
  \end{tikzpicture}%
  }
  \caption[Architecture of tensor-network-editor]{Architecture of \texttt{tensor-network-editor}. Editor sessions, CLI/Python builders, templates, supported imports, and optional limited live imports produce a backend-independent NetworkSpec design model. The model primarily drives backend code generation, while validation, linting, analysis, contraction planning, static export, and versioned JSON preservation support the visual-to-code authoring workflow.}
  \label{fig:architecture-tn-editor}
\end{figure}

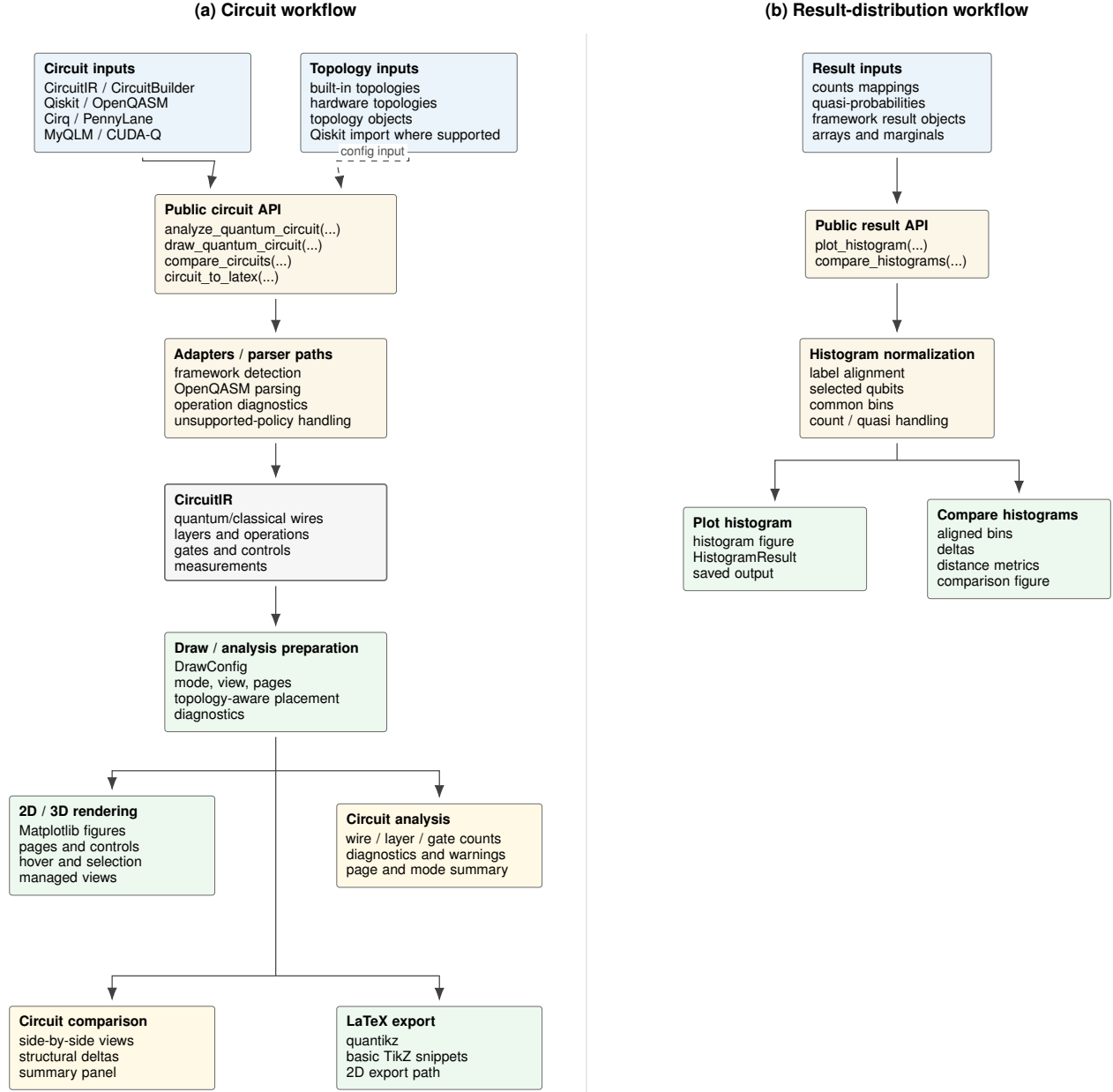
\begin{figure}[!htbp]
  \centering
  \resizebox{\linewidth}{!}{%
  \begin{tikzpicture}[x=1cm,y=1cm]
  \node[font=\sffamily\bfseries\fontsize{9}{9.5}\selectfont] at (4.4,15.75) {(a) Circuit workflow};
  \node[font=\sffamily\bfseries\fontsize{9}{9.5}\selectfont] at (15.6,15.75) {(b) Result-distribution workflow};
  \draw[black!15, line width=0.5pt] (10.0,-3.9) -- (10.0,15.35);

  \node[archbox, fill=archsoftblue, text width=3.55cm] (f3cinputs) at (2.0,14.1)
    {\textbf{Circuit inputs}\\[0.6mm]
    CircuitIR / CircuitBuilder\\
    Qiskit / OpenQASM\\
    Cirq / PennyLane\\
    MyQLM / CUDA-Q};

  \node[archbox, fill=archsoftblue, text width=3.55cm] (f3topology) at (6.8,14.1)
    {\textbf{Topology inputs}\\[0.6mm]
    built-in topologies\\
    hardware topologies\\
    topology objects\\
    Qiskit import where supported};

  \node[archbox, fill=archsofttan, text width=4.0cm] (f3api) at (4.4,11.55)
    {\textbf{Public circuit API}\\[0.6mm]
    analyze\_quantum\_circuit(...)\\
    draw\_quantum\_circuit(...)\\
    compare\_circuits(...)\\
    circuit\_to\_latex(...)};

  \node[archbox, fill=archsofttan, text width=3.65cm] (f3adapters) at (4.4,8.95)
    {\textbf{Adapters / parser paths}\\[0.6mm]
    framework detection\\
    OpenQASM parsing\\
    operation diagnostics\\
    unsupported-policy handling};

  \node[archcorebox, fill=archsoftgray, text width=3.65cm] (f3ir) at (4.4,6.35)
    {\textbf{CircuitIR}\\[0.6mm]
    quantum/classical wires\\
    layers and operations\\
    gates and controls\\
    measurements};

  \node[archbox, fill=archsoftgreen, text width=3.65cm] (f3prep) at (4.4,3.65)
    {\textbf{Draw / analysis preparation}\\[0.6mm]
    DrawConfig\\
    mode, view, pages\\
    topology-aware placement\\
    diagnostics};

  \node[archbox, fill=archsoftgreen, text width=3.35cm] (f3render) at (1.45,0.7)
    {\textbf{2D / 3D rendering}\\[0.6mm]
    Matplotlib figures\\
    pages and controls\\
    hover and selection\\
    managed views};

  \node[archbox, fill=archsoftyellow, text width=3.35cm] (f3analysis) at (7.35,0.7)
    {\textbf{Circuit analysis}\\[0.6mm]
    wire / layer / gate counts\\
    diagnostics and warnings\\
    page and mode summary};

  \node[archbox, fill=archsoftyellow, text width=3.35cm] (f3comparison) at (1.45,-2.95)
    {\textbf{Circuit comparison}\\[0.6mm]
    side-by-side views\\
    structural deltas\\
    summary panel};

  \node[archbox, fill=archsoftgreen, text width=3.35cm] (f3latex) at (7.35,-2.95)
    {\textbf{LaTeX export}\\[0.6mm]
    quantikz\\
    basic TikZ snippets\\
    2D export path};

  \node[archbox, fill=archsoftblue, text width=3.05cm] (f3rinputs) at (15.6,14.1)
    {\textbf{Result inputs}\\[0.6mm]
    counts mappings\\
    quasi-probabilities\\
    framework result objects\\
    arrays and marginals};

  \node[archbox, fill=archsofttan, text width=2.95cm] (f3rapi) at (15.6,11.55)
    {\textbf{Public result API}\\[0.6mm]
    plot\_histogram(...)\\
    compare\_histograms(...)};

  \node[archbox, fill=archsofttan, text width=3.15cm] (f3hnorm) at (15.6,8.95)
    {\textbf{Histogram normalization}\\[0.6mm]
    label alignment\\
    selected qubits\\
    common bins\\
    count / quasi handling};

  \node[archbox, fill=archsoftgreen, text width=2.95cm] (f3plot) at (13.4,6.05)
    {\textbf{Plot histogram}\\[0.6mm]
    histogram figure\\
    HistogramResult\\
    saved output};

  \node[archbox, fill=archsoftgreen, text width=2.95cm] (f3hcompare) at (17.8,6.05)
    {\textbf{Compare histograms}\\[0.6mm]
    aligned bins\\
    deltas\\
    distance metrics\\
    comparison figure};

  \path (f3api.north) ++(-1.2,0) coordinate (f3apiinleft);
  \path (f3api.north) ++(1.2,0) coordinate (f3apiinright);

  \draw[archarrow] (f3cinputs.south) -- (2.0,13.05) -- (3.3,13.05) -- (f3apiinleft);
  \draw[archoptional] (f3topology.south) -- (6.8,13.05) -- node[archarrowlabel, above] {config input} (5.5,13.05) -- (f3apiinright);
  \draw[archarrow] (f3api.south) -- (f3adapters.north);
  \draw[archarrow] (f3adapters.south) -- (f3ir.north);
  \draw[archarrow] (f3ir.south) -- (f3prep.north);
  \draw[archline] (f3prep.south) -- (4.4,-1.65);
  \draw[archbranch] (4.4,2.05) -| (f3render.north);
  \draw[archbranch] (4.4,2.05) -| (f3analysis.north);
  \draw[archbranch] (4.4,-1.65) -| (f3comparison.north);
  \draw[archbranch] (4.4,-1.65) -| (f3latex.north);

  \draw[archarrow] (f3rinputs.south) -- (f3rapi.north);
  \draw[archarrow] (f3rapi.south) -- (f3hnorm.north);
  \draw[archline] (f3hnorm.south) -- (15.6,7.65);
  \draw[archbranch] (15.6,7.65) -| (f3plot.north);
  \draw[archbranch] (15.6,7.65) -| (f3hcompare.north);
  \end{tikzpicture}%
  }
  \caption[Architecture of quantum-circuit-drawer]{Architecture of \texttt{quantum-circuit-drawer}. Panel (a) shows the circuit workflow: supported circuit objects, OpenQASM inputs, topology configuration, and public drawing/analysis calls are normalized through adapter and parser paths into CircuitIR-based preparation for 2D/3D rendering, circuit analysis, circuit comparison, and LaTeX export. Panel (b) shows the sibling result-distribution workflow: counts, quasi-probabilities, arrays, marginals, and supported result objects are normalized for histogram plotting and distribution comparison; this path visualizes and compares results but does not execute circuits.}
  \label{fig:architecture-qc-drawer}
\end{figure}

Figures~\ref{fig:architecture-tn-visualization}--\ref{fig:architecture-qc-drawer}
show the package architectures because their central representations and user
workflows differ: extracted tensor-network graphs, preserved editable design
models, circuit visualization models, and result-distribution models.

\clearpage

\FloatBarrier

\section{Package Overview}

Table~\ref{tab:package-overview} summarizes the three packages.
The rows are phrased in terms of user-facing problems and workflow contributions rather than as performance claims.

\begin{table}[!htbp]
\centering
\caption{Overview of the three packages in user-problem terms. Backend support is conditional on documented adapters, public APIs, and optional dependencies.}
\begingroup
\scriptsize
\setlength{\tabcolsep}{2pt}
\renewcommand{\arraystretch}{1.15}
\begin{tabular}{@{}L{0.16\linewidth} L{0.29\linewidth} L{0.25\linewidth} L{0.24\linewidth}@{}}
\toprule
Package & Main user problem & Core contribution & Typical use \\
\midrule
\toolname{Tensor-Network-Visualization} & ``I already have a tensor network or \code{einsum} workflow in code, but I need to debug and analyze whether it is connected and contracted as intended.'' & Visual debugging of network structure and tensor contents, with selected translations as complementary aids for supported inputs & Debugging construction, checking contractions, inspecting intermediate or resulting tensors \\
\toolname{Tensor-Network-Editor} & ``I want to create tensor-network code more easily, with fewer structural mistakes, especially for a complex or non-standard network.'' & Visual-to-code authoring and backend code generation, with preservation, export, and analysis as supporting features & Building a network visually, generating backend code, and then exporting or preserving the resulting design \\
\toolname{Quantum Circuit Drawer} & ``I need to visualize a circuit clearly and analyze it comfortably.'' & Clear circuit rendering and inspection-oriented analysis, with circuit or result comparison as complementary workflows & Debugging circuits, teaching, communicating circuit structure, and inspecting result distributions when needed \\
\bottomrule
\end{tabular}
\endgroup
\label{tab:package-overview}
\end{table}

\toolname{Tensor-Network-Visualization} is distributed as \code{tensor-network-visualization} and imported as \code{tensor\_network\_viz}.
The version considered here is 2.0.3, with MIT licensing, Python 3.11 or newer, and base dependencies on NumPy, Matplotlib, and NetworkX.
Documented inputs include TensorKrowch, TensorNetwork, Quimb, explicit TeNPy-style structures, and traced NumPy/PyTorch \code{einsum} workflows, through documented adapter paths.
Public entry points include \code{show\_tensor\_network(...)}, \code{show\_tensor\_elements(...)}, \code{show\_tensor\_comparison(...)}, \code{translate\_tensor\_network(...)}, \code{normalize\_tensor\_network(...)}, and \code{export\_tensor\_network\_snapshot(...)}.
Its primary emphasis in this manuscript is visual debugging and analysis of already programmed or generated tensor networks: seeing whether a network is wired as expected, examining contraction-oriented structure, and inspecting selected tensor values.
Translation and comparison utilities are complementary aids for migration or side-by-side checks on supported inputs.

\toolname{Tensor-Network-Editor} is distributed as \code{tensor-network-editor} and imported as \code{tensor\_network\_editor}.
The version considered here is 1.0.1, with MIT licensing, Python 3.11 or newer, and runtime dependencies including Matplotlib and \code{opt\_einsum}.
The package provides a local browser-served editor, JSON design files, selected backend code generation, and static exports including SVG, PNG, PDF, TikZ/LaTeX, Graphviz/DOT, and Mermaid.
Its primary role here is to help users generate tensor-network code more easily from a visual design, especially when the intended network is custom, non-standard, or awkward to assemble directly in code.
Visual editing, JSON preservation, export, and design-level analysis are useful supporting features, but they are secondary to that visual-to-code workflow.

\toolname{Quantum Circuit Drawer} is distributed as \code{quantum-circuit-drawer} and imported as \code{quantum\_circuit\_drawer}.
The version considered here is 1.1.1, with MIT licensing, Python 3.11 or newer, and base dependencies on Matplotlib and NumPy.
Its documented workflows include \code{analyze\_quantum\_circuit(...)}, \code{draw\_quantum\_circuit(...)}, \code{circuit\_to\_latex(...)}, \code{compare\_circuits(...)}, \code{plot\_histogram(...)}, and \code{compare\_histograms(...)}.
Documented adapter paths include the package's internal circuit representation, Qiskit, OpenQASM 2, OpenQASM 3, Cirq, PennyLane, MyQLM, and CUDA-Q, with installation extras and platform-qualified adapter paths summarized below.
In this paper, the package's primary role is to render circuits clearly and make them easy to inspect and analyze during development, teaching, or communication.
Comparison of circuits or result distributions is useful, but it is treated here as complementary to that main visualization-and-analysis role.
Where supported by the package and input type, the inspection workflow can include managed views for larger circuits, decomposition-oriented inspection, and topology-aware layouts.

\subsection{Installation and optional dependencies}

All three packages require Python 3.11 or newer. Their base installations are intentionally kept separate from heavier framework integrations: the core packages support the documented lightweight rendering, editing, inspection, or data-model workflows, while backend-specific imports, tracing, notebook interaction, desktop wrapping, or quantum-framework adapters are enabled through optional extras.

\begin{table}[t]
\centering
\caption{Base installation and optional extras for the three packages.\protect\footnotemark[2]}
\scriptsize
\setlength{\tabcolsep}{2pt}
\renewcommand{\arraystretch}{1.15}
\begin{tabularx}{\textwidth}{@{}L{0.14\textwidth} L{0.19\textwidth} L{0.24\textwidth} X@{}}
\toprule
Package & Base install & Base runtime role & Extras list \\
\midrule
\toolname{Tensor-Network-Visualization} &
\texttt{python -m pip install}\newline
\texttt{tensor-network-}\newline
\texttt{visualization} &
NumPy, Matplotlib, and NetworkX based visualization and normalized inspection paths. &
\texttt{jupyter}, \texttt{tensorkrowch}, \texttt{tensornetwork}, \texttt{quimb}, \texttt{tenpy}, \texttt{einsum} \\
\toolname{Tensor-Network-Editor} &
\texttt{python -m pip install}\newline
\texttt{tensor-network-}\newline
\texttt{editor} &
Local visual editor, JSON design model, rendering/export paths, and code-generation infrastructure. &
\texttt{numpy}, \texttt{torch}, \texttt{tensornetwork}, \texttt{quimb}, \texttt{tensorkrowch}, \texttt{desktop} \\
\toolname{Quantum Circuit Drawer} &
\texttt{python -m pip install}\newline
\texttt{quantum-circuit-}\newline
\texttt{drawer} &
Matplotlib/NumPy based circuit drawing, internal IR, histogram plotting, and documented core rendering paths. &
\texttt{qiskit}, \texttt{qasm3}, \texttt{cirq}, \texttt{pennylane}, \texttt{myqlm}, \texttt{cudaq}, \texttt{notebook} \\
\bottomrule
\end{tabularx}
\label{tab:installation-extras}
\end{table}
\footnotetext[2]{\texttt{tenpy} resolves to \texttt{physics-tenpy}. \texttt{einsum} adds PyTorch tracing support. \texttt{desktop} refers to \texttt{pywebview} desktop mode. \texttt{qasm3} depends on Qiskit plus \texttt{qiskit-qasm3-import}. \texttt{cudaq} is Linux/WSL2 only. \texttt{notebook} adds Jupyter interaction support.}

Consequently, examples in this manuscript assume that the corresponding optional extras have been installed when a backend-specific adapter, traced einsum path, notebook interaction mode, desktop wrapper, or quantum-framework import path is used.
\FloatBarrier

\section{Reproducible Examples}

\subsection{Tensor-network visualization}

\begin{verbatim}
import numpy as np
from tensor_network_viz import EinsumTrace, PlotConfig, einsum, show_tensor_network

trace = EinsumTrace()
a = np.ones((2, 3), dtype=float)
x = np.array([1.0, -0.5, 0.25], dtype=float)
trace.bind("A", a)
trace.bind("x", x)
einsum("ab,b->a", a, x, trace=trace, backend="numpy")

fig, ax = show_tensor_network(
    trace,
    config=PlotConfig(show_tensor_labels=True, hover_labels=True),
    show=False,
)
fig.savefig("einsum-network.png", bbox_inches="tight")
\end{verbatim}

This example is intended as a visual debugging workflow rather than as a numerical benchmark.
The goal is to make a traced \code{einsum} network visible so that connectivity, visible indices, or contraction structure can be checked before or after computation.

\begin{figure}[!htbp]
  \centering
  \setlength{\tabcolsep}{3pt}
  \renewcommand{\arraystretch}{0}
  \begin{tabular}{cc}
    \includegraphics[width=0.47\linewidth]{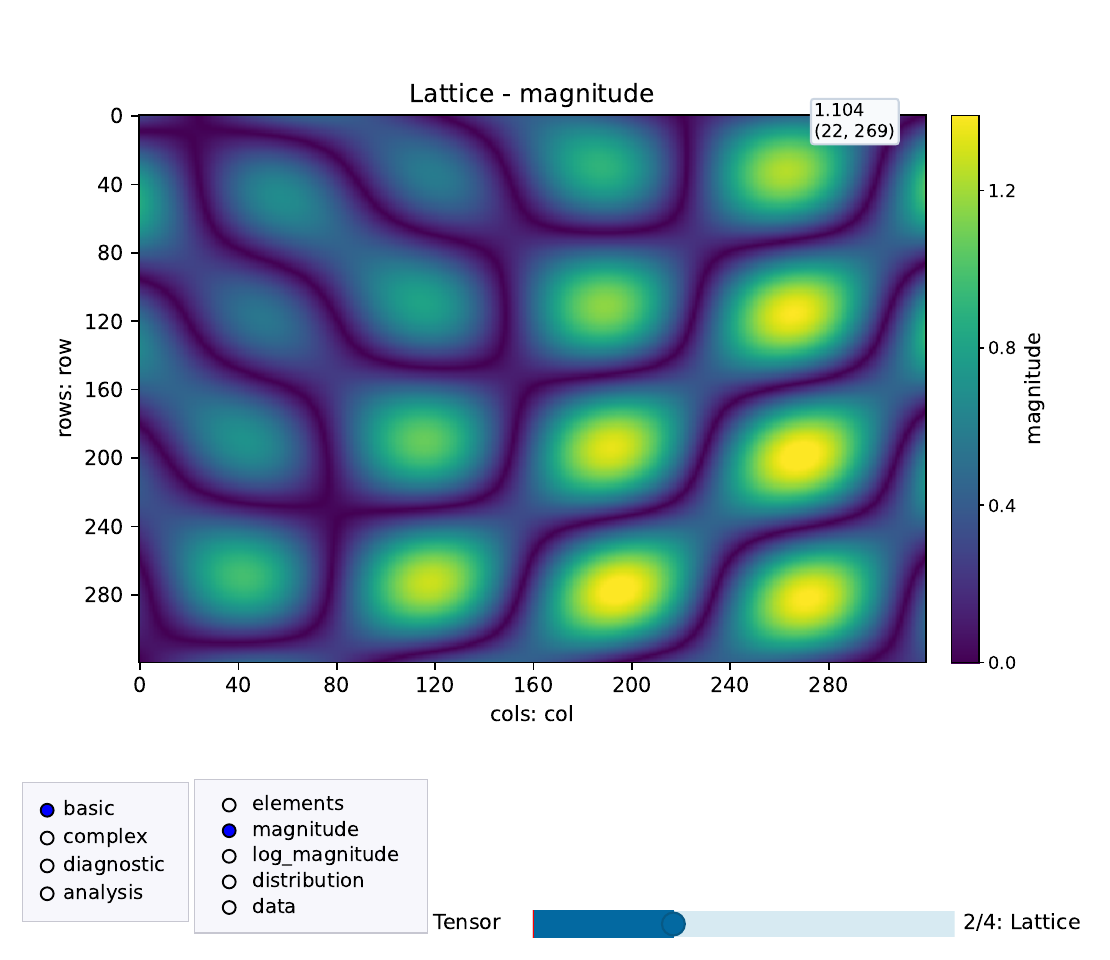} &
    \includegraphics[width=0.47\linewidth]{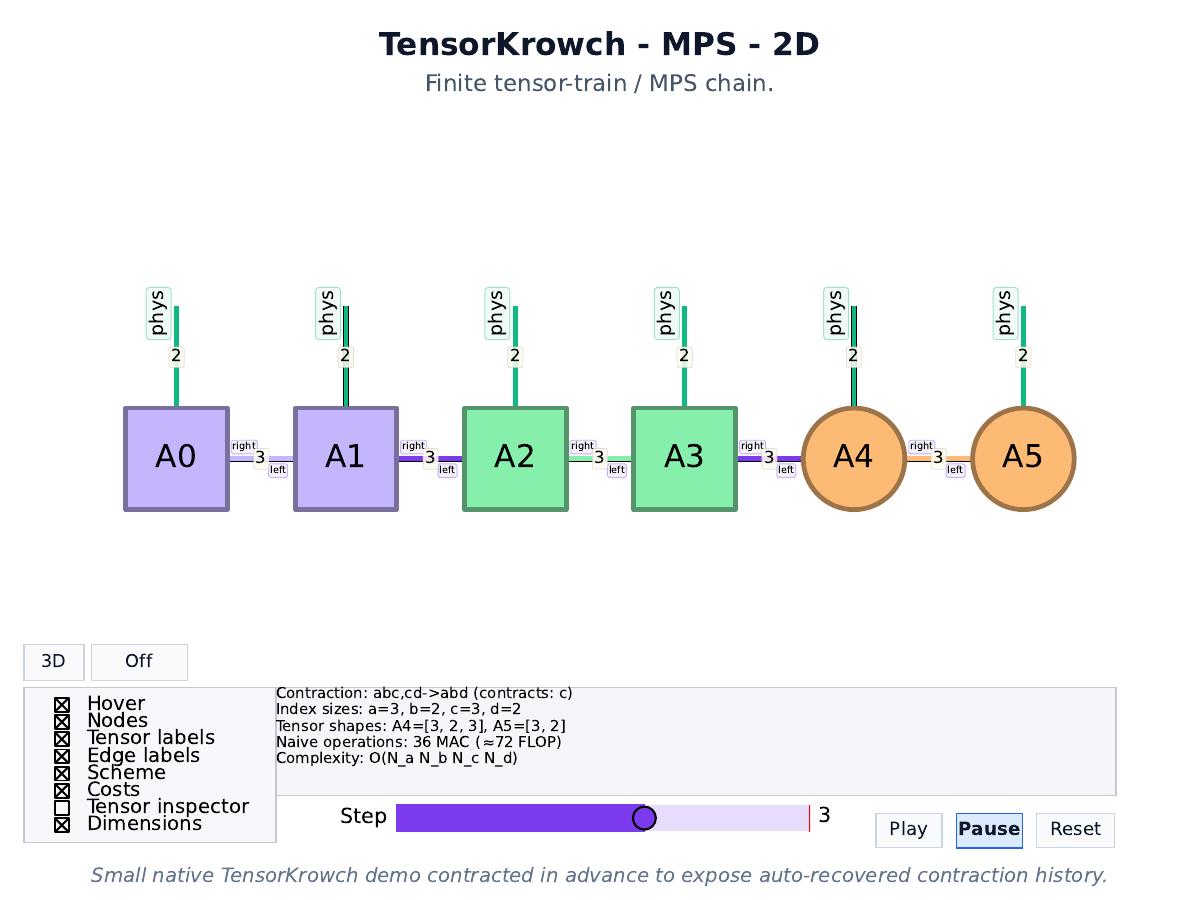} \\
    \includegraphics[width=0.47\linewidth]{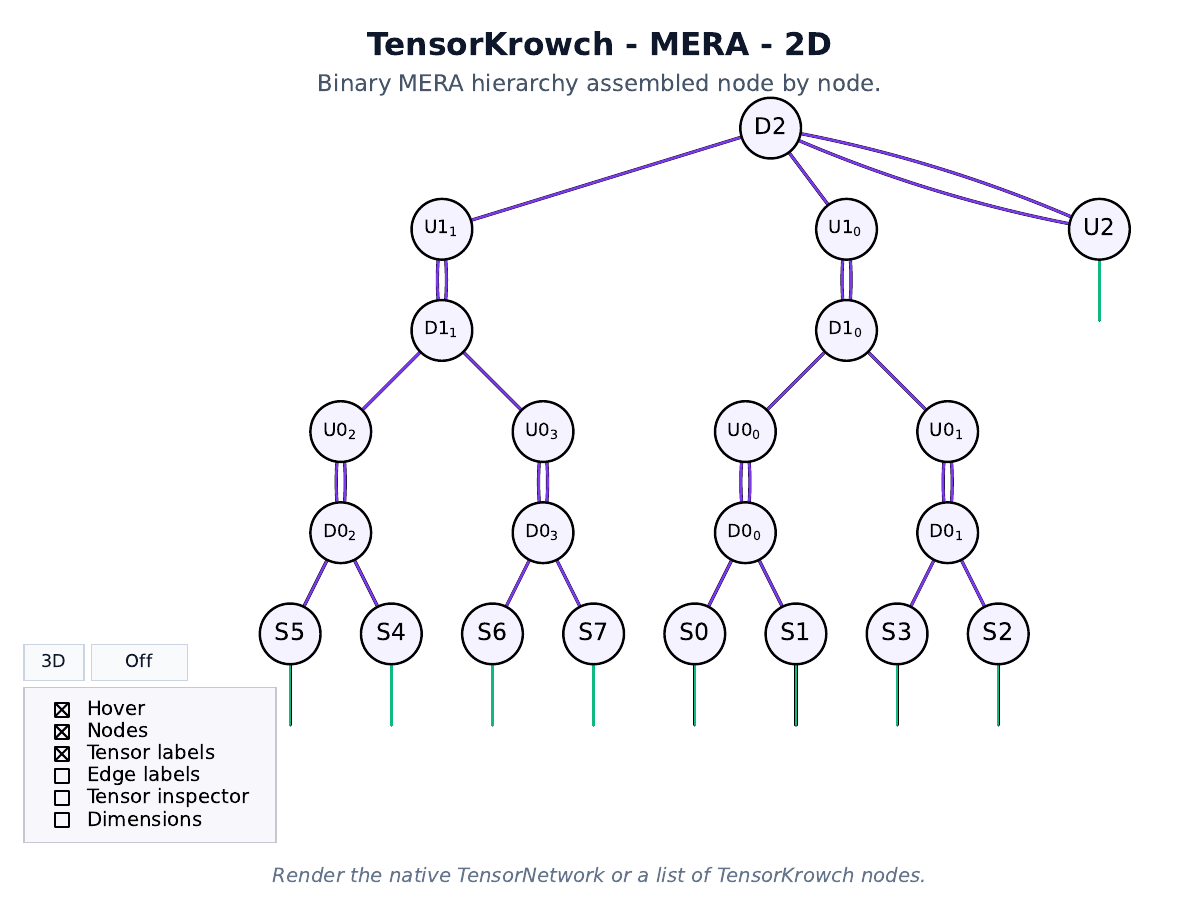} &
    \includegraphics[width=0.47\linewidth]{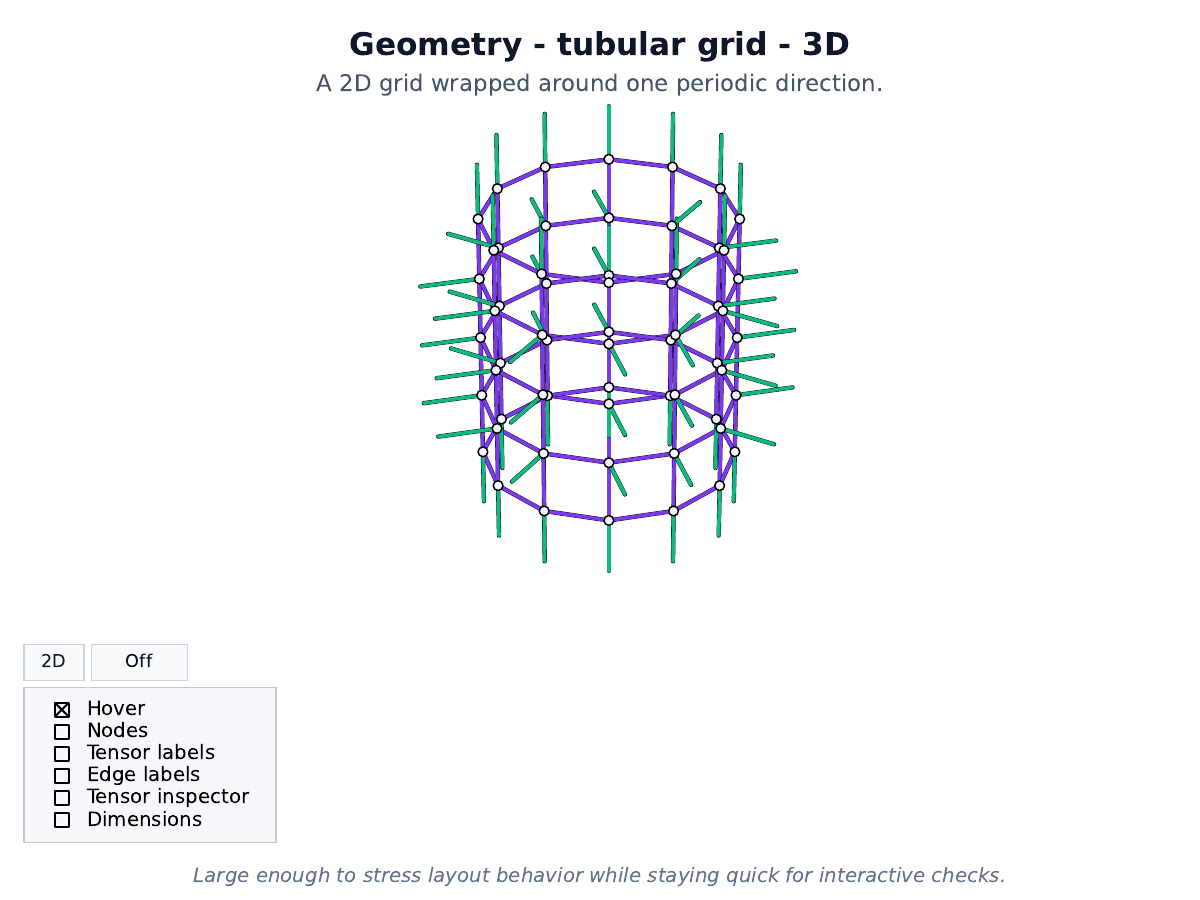}
  \end{tabular}
  \caption{Representative views from the media set associated with \texttt{tensor-network-visualization} version 2.0.3: an inspection-oriented network view, an MPS and contraction-oriented example, a MERA-style layout, and a tubular rendering, illustrating the package's range from structural debugging to richer geometry-aware inspection.}
  \label{fig:tnviz-gallery}
\end{figure}
\FloatBarrier

\subsection{Tensor-network code generation from a design specification}

The following example uses the documented programmatic builder rather than a browser session so that the manuscript example remains directly executable. It exercises the same \code{NetworkSpec}-centered code-generation path used by saved visual designs: the network structure is first represented as a backend-independent design specification and is then passed to the selected code-generation target.

\begin{verbatim}
from tensor_network_editor import EngineName, NetworkBuilder, generate_code

builder = NetworkBuilder("demo")
a = builder.tensor("A")
a.index("i", 2)
a.index("x", 3)
b = builder.tensor("B")
b.index("x", 3)
b.index("j", 4)
builder.connect(a["x"], b["x"], name="bond_x")
spec = builder.build()

result = generate_code(spec, engine=EngineName.EINSUM_NUMPY)
print(result.code)
\end{verbatim}

This two-tensor builder example follows the documented Python path and isolates the code-generation layer from the graphical user interface. It should therefore be read as a reproducible check of the design-to-code pipeline rather than as a screenshot-level demonstration of the local editor. In the full visual workflow, the same kind of design specification can be produced through the editor, preserved as JSON, and passed through the documented generation and export paths. Templates, imports, planners, and render/export commands remain part of the broader workflow described in the package documentation.

\begin{figure}[!htbp]
  \centering
  \setlength{\tabcolsep}{3pt}
  \renewcommand{\arraystretch}{0}
  \begin{tabular}{cc}
    \includegraphics[width=0.47\linewidth]{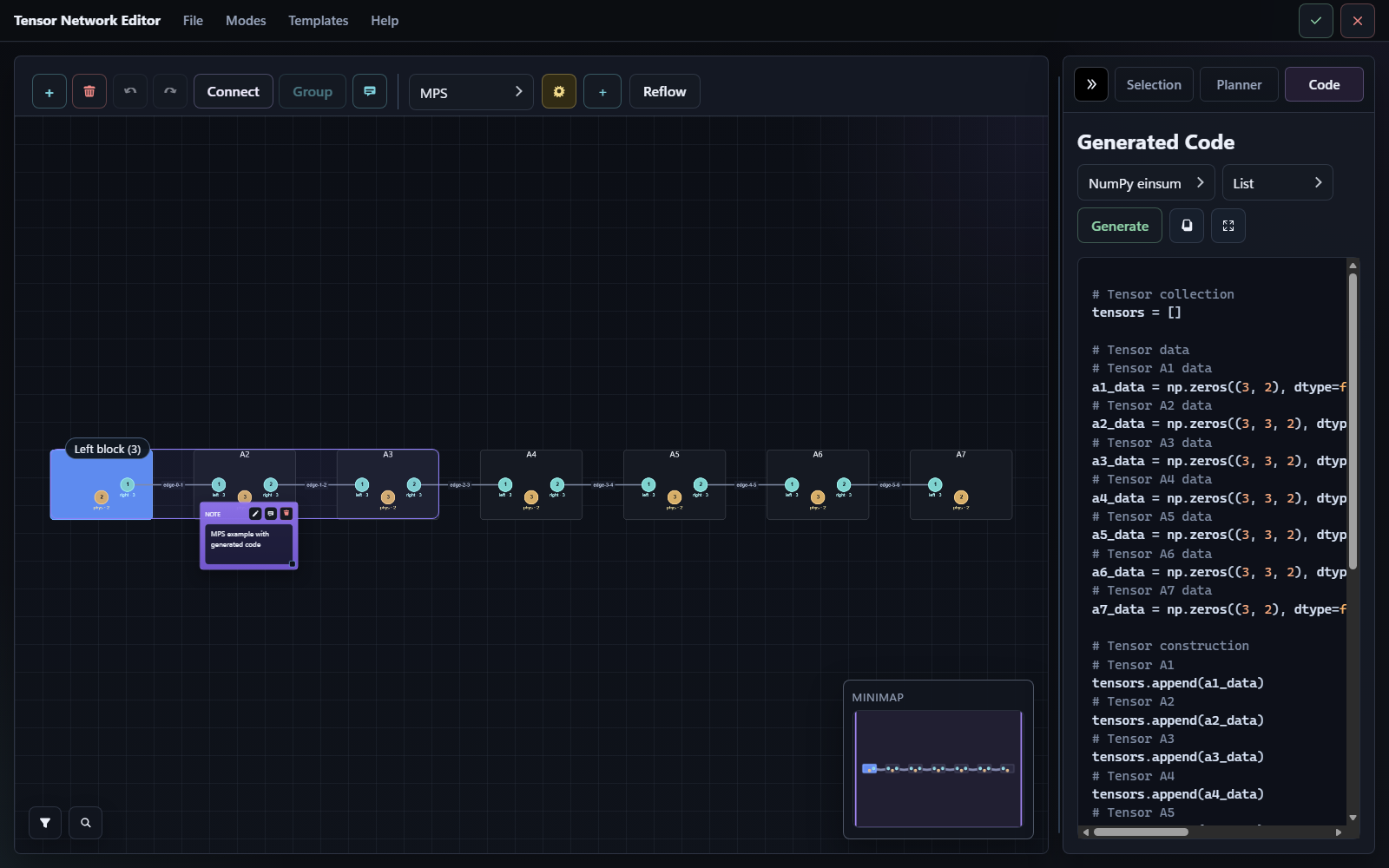} &
    \includegraphics[width=0.47\linewidth]{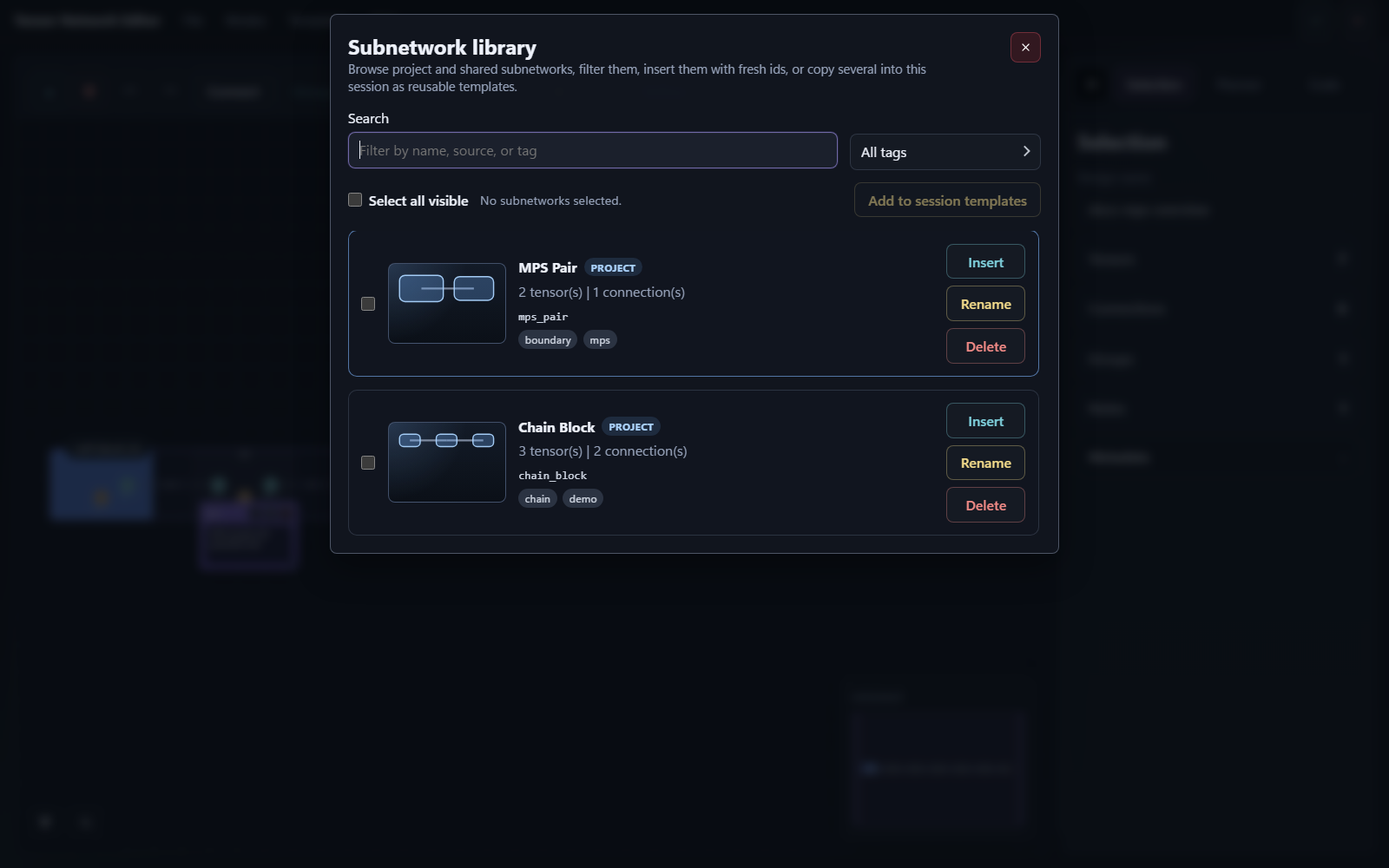} \\
    \includegraphics[width=0.47\linewidth]{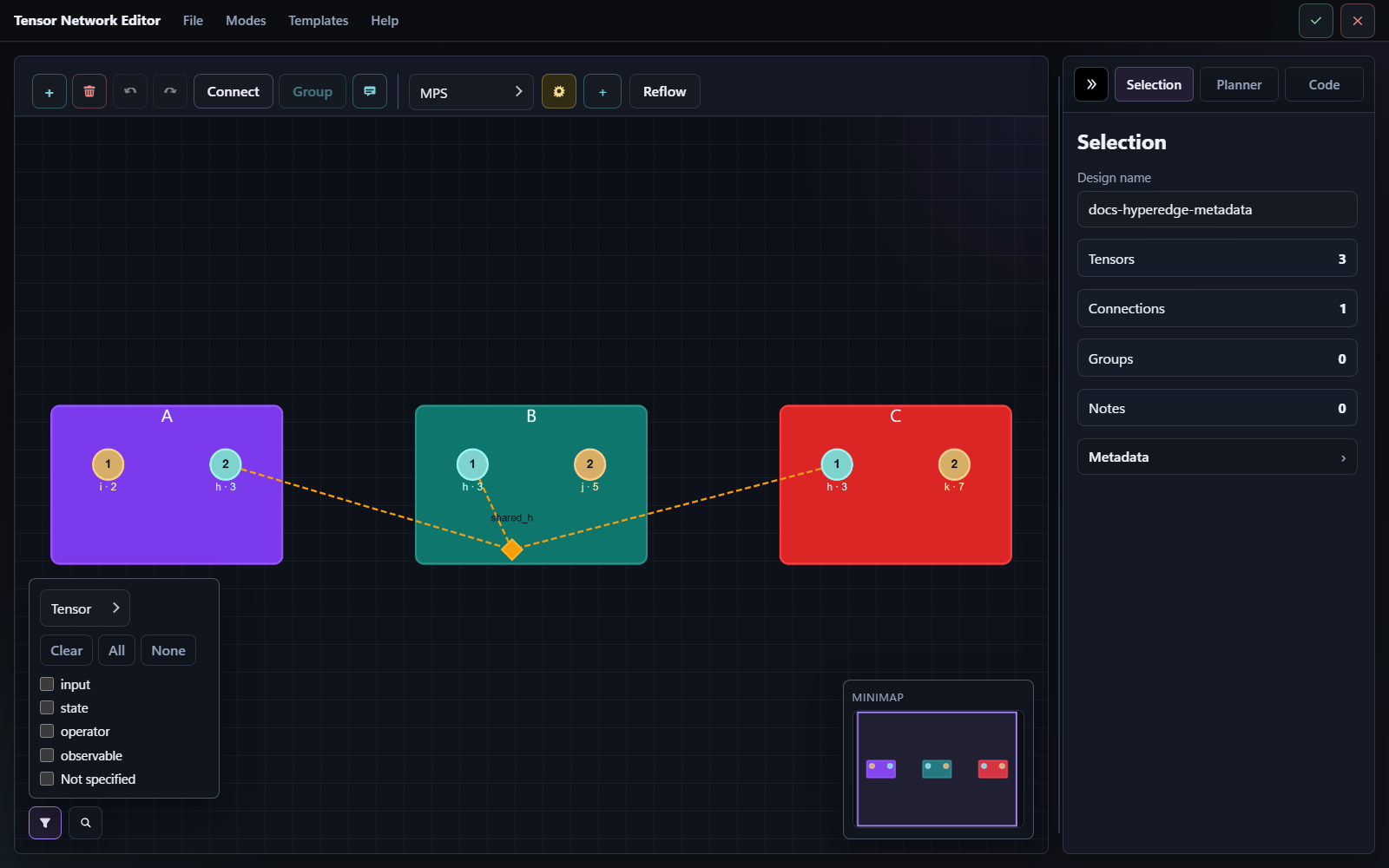} &
    \includegraphics[width=0.47\linewidth]{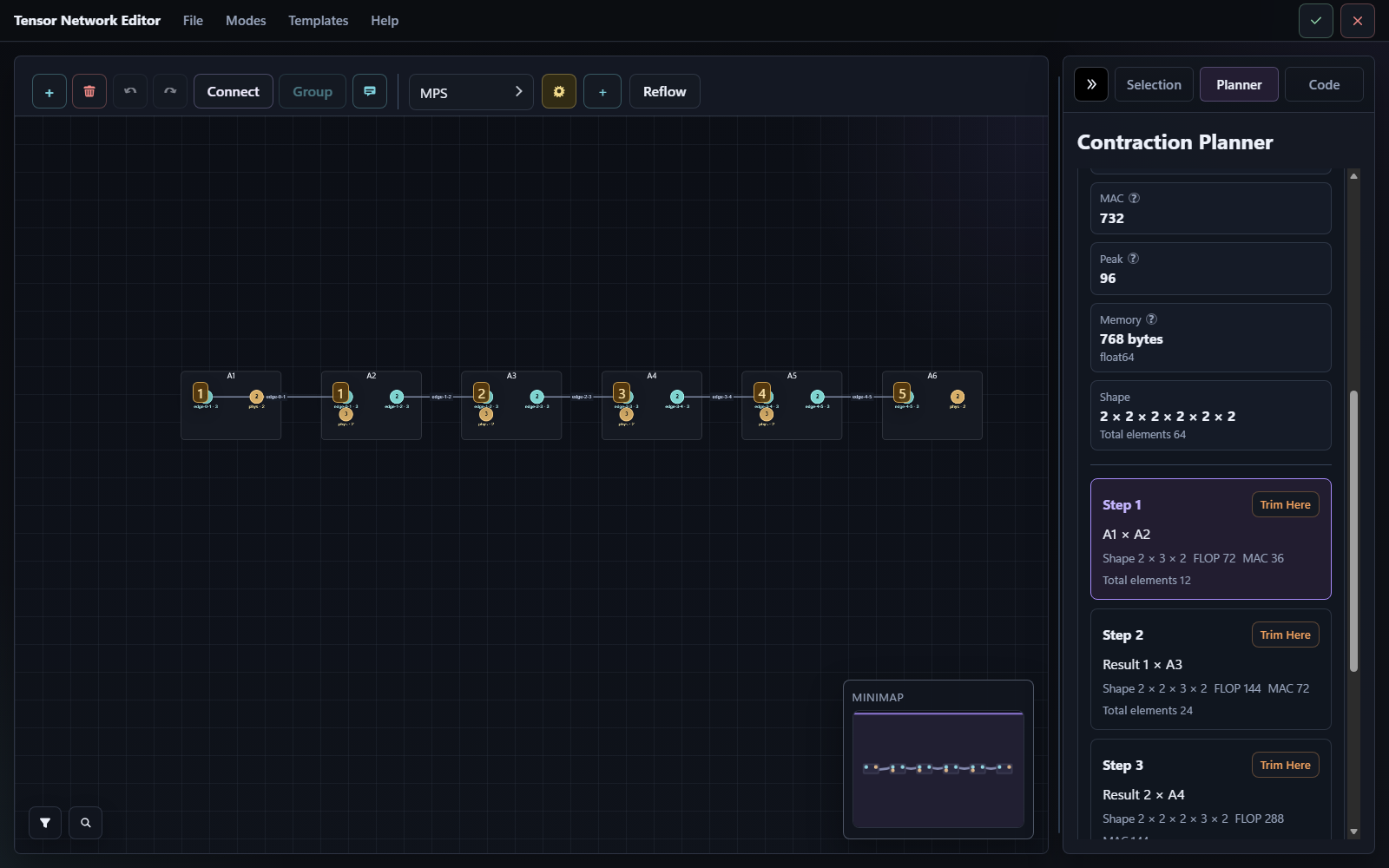}
  \end{tabular}
  \caption{Representative local-editor views from the \texttt{tensor-network-editor} README gallery for version 1.0.1: the main authoring canvas with generated code, template and reusable-subnetwork tooling, hyperedge and metadata editing, and the manual contraction planner. The gallery illustrates the interactive visual authoring interface; the executable example in the text isolates the shared design-to-code path.}
  \label{fig:tne-editor-gallery}
\end{figure}
\FloatBarrier

\subsection{Quantum-circuit rendering and inspection}

\begin{verbatim}
from qiskit import QuantumCircuit

from quantum_circuit_drawer import draw_quantum_circuit

circuit = QuantumCircuit(2, 2)
circuit.h(0)
circuit.cx(0, 1)
circuit.measure([0, 1], [0, 1])

draw_quantum_circuit(
    circuit,
    output_path="bell.png",
    show=False,
)
\end{verbatim}

This minimal example illustrates the common rendering path highlighted in the README quick-start documentation for version 1.1.1.
The code constructs a Bell-state circuit in Qiskit only as an input object for drawing: the package renders and inspects the circuit structure but does not execute, simulate, or validate it.
In the intended workflow, such rendering supports clear inspection and analysis during explanation, debugging, and communication.

\begin{figure}[!htbp]
  \centering
  \setlength{\tabcolsep}{3pt}
  \renewcommand{\arraystretch}{0}
  \begin{tabular}{cc}
    \includegraphics[width=0.47\linewidth]{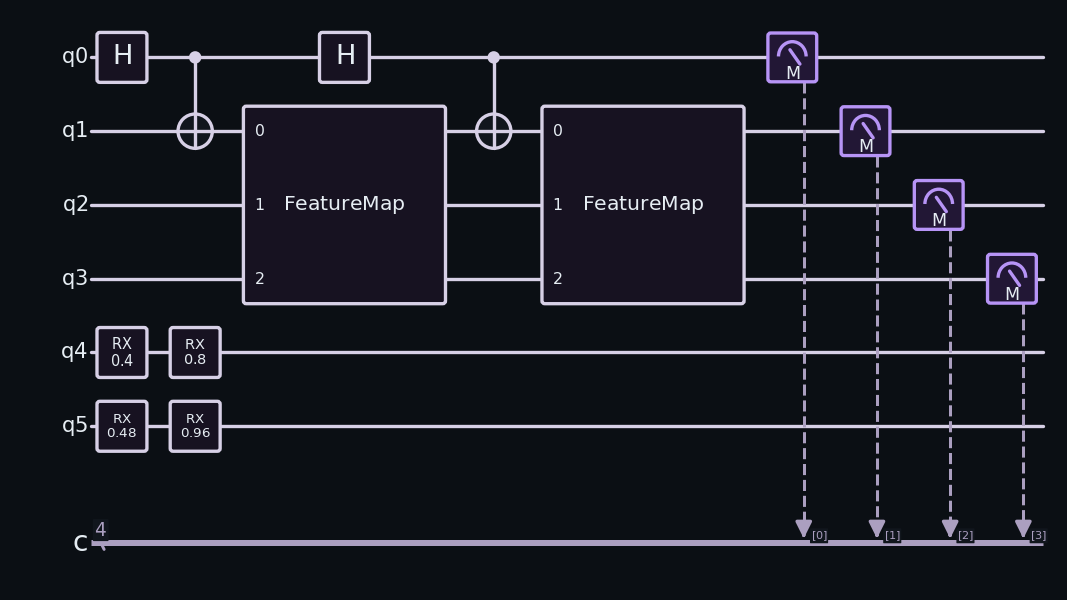} &
    \includegraphics[width=0.47\linewidth]{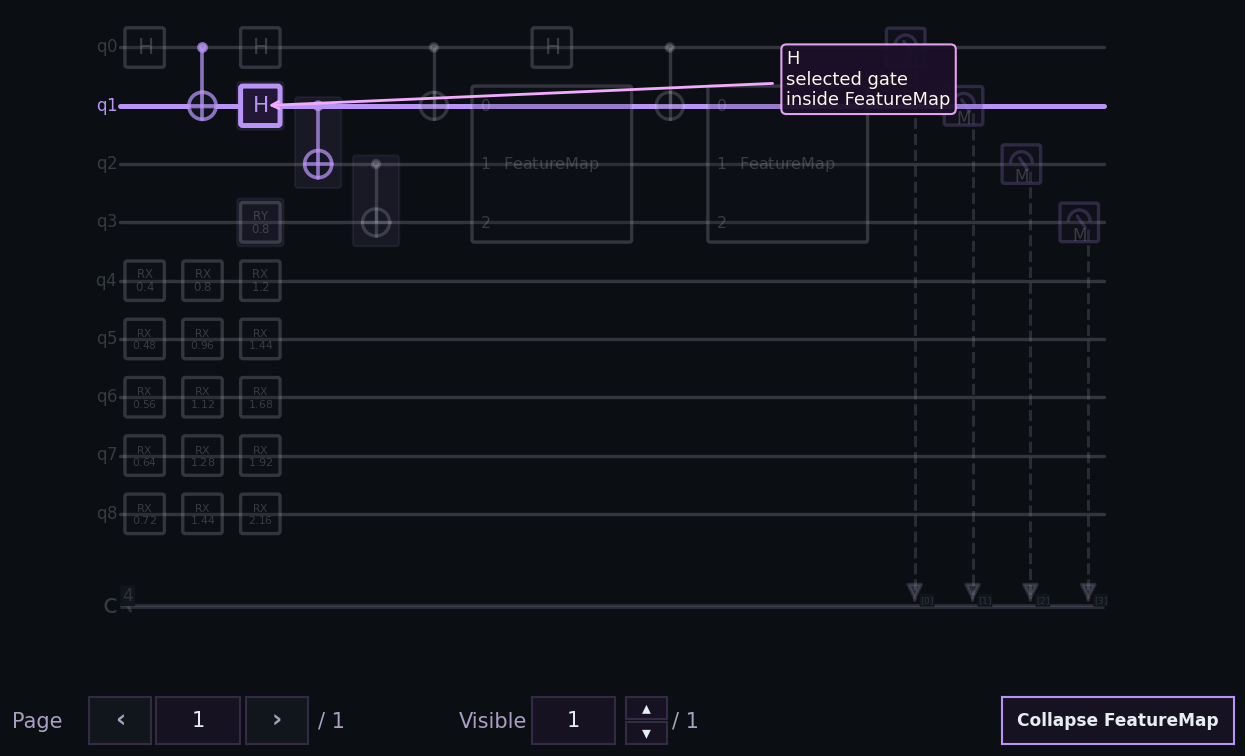} \\
    \includegraphics[width=0.47\linewidth]{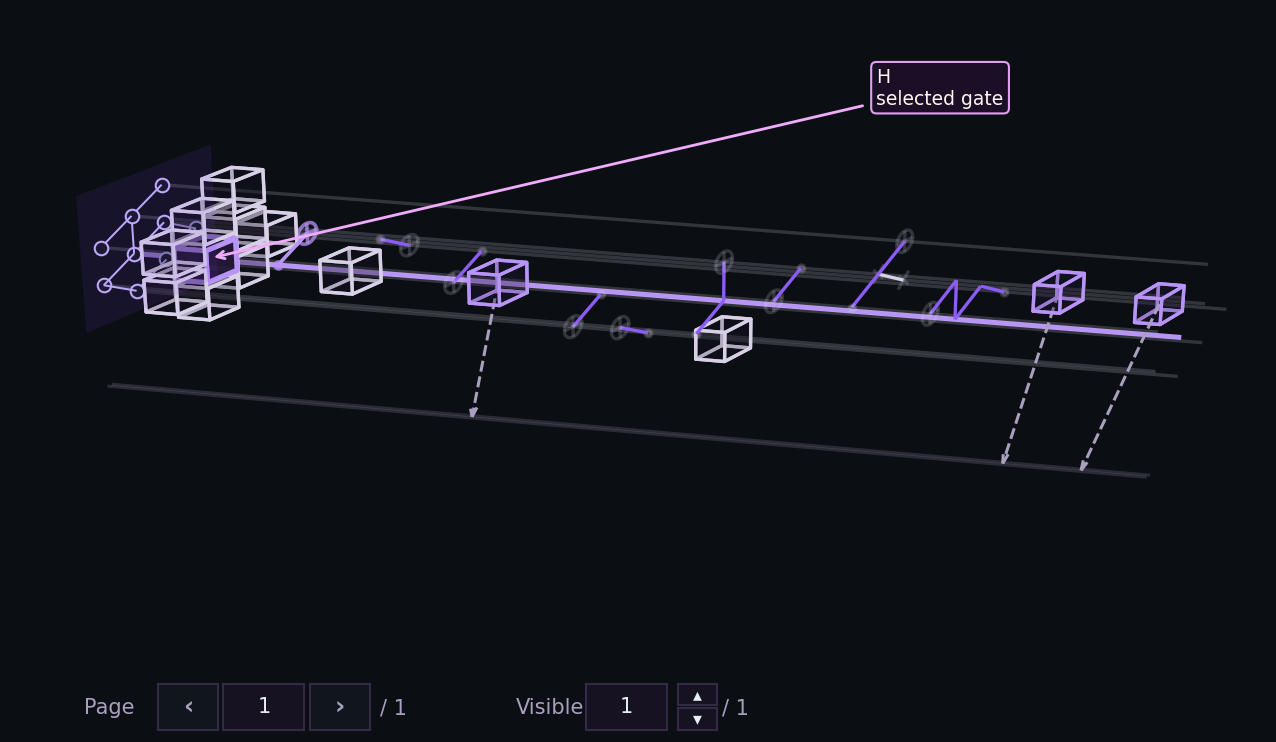} &
    \includegraphics[width=0.42\linewidth]{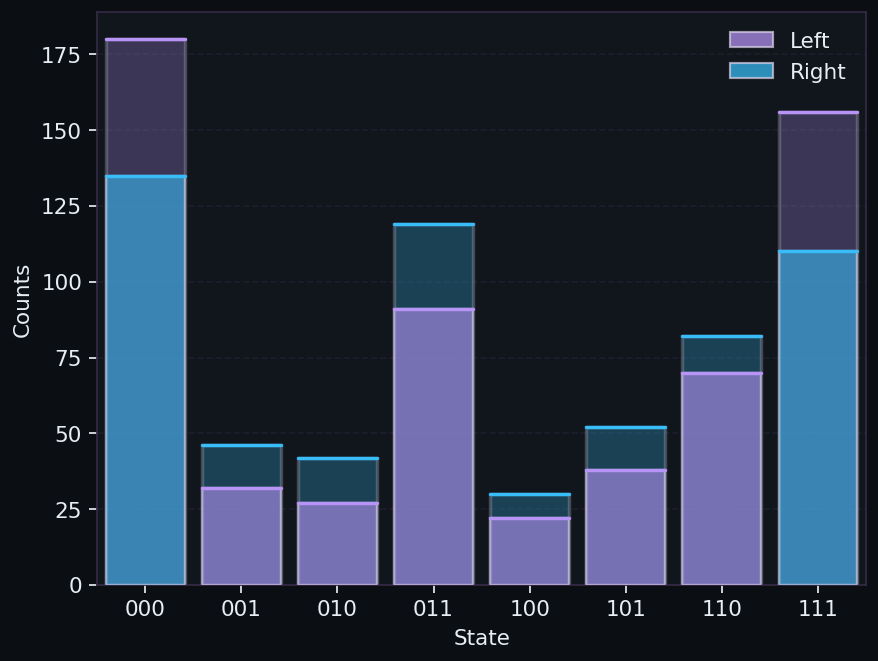}
  \end{tabular}
  \caption{Representative views from the \texttt{quantum-circuit-drawer} README gallery for version 1.1.1: a static 2D circuit, a managed 2D exploration view, a topology-aware 3D rendering, and a result-distribution comparison, illustrating the package's range from circuit inspection to complementary output analysis.}
  \label{fig:qcd-gallery}
\end{figure}
\FloatBarrier

\subsection{Complementary result-distribution comparison}

\begin{verbatim}
from quantum_circuit_drawer import compare_histograms

ideal = {"00": 0.5, "11": 0.5}
sampled = {"00": 473, "01": 19, "10": 24, "11": 484}

result = compare_histograms(
    ideal,
    sampled,
    sort="delta_desc",
    left_label="Ideal",
    right_label="Sampled",
    show=False,
)
\end{verbatim}

This example focuses on a complementary analysis workflow: visual comparison of result distributions rather than execution.
The ideal distribution matches the two-qubit correlations expected from a Bell-like experiment, while the sampled counts illustrate how two documented outputs can be compared side by side.
The package plots and compares these distributions for inspection and communication; it does not execute circuits or produce the counts itself.
The intended use is to inspect whether two outputs differ in a way that is easy to communicate, debug, or compare across runs or frameworks.
\FloatBarrier

\section{Scope, Limitations, and Preservation Guarantees}

The scope contract is intentionally conservative.
The packages preserve the structural information needed for their documented visualization, authoring, export, inspection, and comparison workflows.
They do not guarantee full semantic equivalence for arbitrary backend objects, custom operations, optimizer state, execution metadata, or hardware calibration information.
They also do not claim arbitrary backend preservation, numerical-performance improvements, simulation or compilation capability, or statistical validation of hardware experiments.
The package-specific limitations below make this scope explicit before the preservation summary in Table~\ref{tab:preservation-scope}.

\subsection{Tensor-Network-Visualization}

For TENSOR-NETWORK-VISUALIZATION, the relevant boundary is that normalized tensor graphs and tensor-inspection views are derived from documented inputs and adapter paths. The package does not implement new contraction algorithms, replace backend simulators, or guarantee reconstruction of backend-specific optimizer state, execution internals, custom metadata, or full object semantics. Translation utilities should be treated as migration or comparison aids for supported subsets rather than as universal backend-to-backend converters. Tensor values, labels, shapes, and contraction-oriented metadata are preserved only when they are exposed by the input path or supplied explicitly.

\subsection{Tensor-Network-Editor}

For TENSOR-NETWORK-EDITOR, the strongest guarantees concern the package's own visual design model, versioned JSON records, and documented code-generation targets.
Several boundaries are intentional.
TenPy code generation is out of scope; symbolic tensor expressions remain limited to the portable initializer and data model; and TensorKrowch support is restricted to the documented feasible subset.
Hyperedges are supported only in normal mode and are lowered to generated copy tensors for export.
Python import is conservative: it supports the package's own generated exports plus static AST patterns for simple Quimb, TensorNetwork, and einsum / opt\_einsum sources.
Live import is available only as a trusted-code workflow for selected Quimb and TensorNetwork runtime objects; it executes local Python files in a subprocess with the active Python environment and does not constitute a security sandbox.
External and live imports do not recover editor layout, groups, notes, or manual contraction plans, and \texttt{PythonLoadOptions(reconstruction\_level="best\_available")} is limited to the package's own generated Python profile.
External static profiles and live imports therefore use the portable \texttt{simple} reconstruction contract.
Manual outer-product steps also remain unsafe to export to TensorKrowch and are treated as a stable project boundary.

\subsection{Quantum Circuit Drawer}

For QUANTUM CIRCUIT DRAWER, the scope boundary is circuit and result-artifact inspection. The platform-qualified adapter paths summarized in Section~5 define the supported import scope; the limitations here concern semantic preservation and execution boundaries. The package renders, analyzes, compares, and exports documented circuit or histogram inputs, but it does not execute circuits, simulate circuits, compile or transpile programs, validate hardware, calibrate hardware, or preserve every framework-specific operation, calibration record, noise model, or execution option. Adapter support is scoped: the internal circuit representation and Qiskit are the strongest documented paths; OpenQASM 2 uses the Qiskit extra; OpenQASM 3 depends on the Qiskit path plus \texttt{qiskit-qasm3-import} where available; Cirq and PennyLane are documented as best-effort on native Windows, with Linux or WSL preferred for repeated reliability; MyQLM is a scoped adapter path when available; and CUDA-Q is limited to Linux or WSL2 and to the documented supported subset. Histogram comparison is an inspection and communication aid, not statistical validation of hardware experiments.

\subsection{Preservation summary}

Table~\ref{tab:preservation-scope} summarizes these boundaries by input or artifact. It is a preservation summary for visualization, authoring, export, inspection, and comparison workflows, not a semantic-equivalence matrix.

\begin{table}[!htbp]
\centering
\caption{Scope and preservation guarantees for representative supported inputs. The entries describe structural information used for visualization, inspection, export, or comparison; they do not imply full semantic equivalence across arbitrary backend-specific objects.}
\begingroup
\scriptsize
\setlength{\tabcolsep}{2pt}
\renewcommand{\arraystretch}{1.15}
\begin{tabular}{@{}L{0.20\linewidth} L{0.25\linewidth} L{0.25\linewidth} L{0.22\linewidth}@{}}
\toprule
Input or artifact & Preserved & Partially preserved & Out of scope \\
\midrule
Quimb tensor networks & Graph topology, visible tensor labels, selected index information & Layout hints, backend-specific metadata & Contraction optimizer state, full backend semantics \\
TensorNetwork, TensorKrowch, and explicit TeNPy inputs & Structural graph information and visible dimensions where available & Tensor values, labels, and custom metadata depending on object path & Backend execution internals and full model semantics \\
TENSOR-NETWORK-EDITOR JSON visual designs & Nodes, edges, labels, layout, and user-visible design metadata & Backend-specific code-generation intent & Guarantees of numerical equivalence after manual edits \\
TENSOR-NETWORK-EDITOR live Python imports & Supported structural information recovered from trusted local Python files through documented import paths & Tensor values, metadata, layout information, groups, notes, or manual plans depending on the import path and object representation & Security sandboxing, safe execution of untrusted code, and guarantees that imported Python code has no side effects \\
OpenQASM strings & Gate sequence and qubit structure for supported syntax & Parameters and custom gates depending on parser support & Hardware calibration and execution backend metadata \\
Qiskit, Cirq, PennyLane, MyQLM, and CUDA-Q inputs & Supported gate and wire structure & Custom operations, parameters, and classical metadata depending on adapter support & Hardware execution, transpilation semantics, and noise models \\
QUANTUM CIRCUIT DRAWER result dictionaries or histograms & Outcome labels and counts or probabilities & Normalization and formatting choices & Statistical validation of hardware experiments \\
\bottomrule
\end{tabular}
\endgroup
\label{tab:preservation-scope}
\end{table}

\section{Research Impact and Use Cases}

The expected impact is primarily practical: the packages target earlier detection of structural errors and clearer communication of tensor-network and circuit artifacts.

\subsection{Debugging traced einsum and programmed tensor networks}
In traced-\code{einsum} or already programmed tensor-network workflows, the practical use case is early structural debugging. A user can inspect whether visible indices, graph connectivity, contraction-oriented views, and selected tensor values match the intended construction before treating the surrounding numerical workflow as final.

\subsection{Visual authoring of non-standard tensor networks}
For custom or non-standard tensor networks, the relevant use case is authoring rather than simulation. A design can be created visually, preserved as JSON, checked at the design level, and then exported as backend code, reducing the amount of hand-written structural boilerplate required before numerical experimentation begins.

\subsection{Inspection of decomposed or transformed quantum circuits}
For quantum-circuit workflows, the use case is inspection of circuits whose structure is difficult to read from code alone, especially after decomposition, import, or framework-specific transformation. Rendered views, managed inspection modes, topology-aware layouts where supported, and distribution-comparison plots help users communicate and compare documented artifacts.

These use cases position the packages as development, teaching, and communication tools. Their value is strongest before a workflow is finalized: when a structure must be checked, explained, exported, or compared in a reproducible visual form.
\FloatBarrier

\section{Availability}

Table~\ref{tab:availability} records the PyPI release date, source repository, and PyPI project page for each package version described in this paper.
PyPI source-distribution SHA256 digests are given in the table footnote.

\begin{table}[!htbp]
\centering
\caption{Release-identification metadata for the package versions described in this paper.\protect\footnotemark[3]}
\begingroup
\scriptsize
\setlength{\tabcolsep}{2pt}
\renewcommand{\arraystretch}{1.15}
\begin{tabular}{@{}L{0.17\linewidth} L{0.07\linewidth} L{0.09\linewidth} L{0.30\linewidth} L{0.30\linewidth}@{}}
\toprule
Package & Version & PyPI release & Source repository & PyPI project \\
\midrule
\shortstack[l]{\texttt{tensor-network-}\\\texttt{visualization}} & 2.0.3 & 17 May 2026 & \url{https://github.com/DOKOS-TAYOS/Tensor-Network-Visualization} & \url{https://pypi.org/project/tensor-network-visualization/} \\
\shortstack[l]{\texttt{tensor-network-}\\\texttt{editor}} & 1.0.1 & 14 May 2026 & \url{https://github.com/DOKOS-TAYOS/Tensor-Network-Editor} & \url{https://pypi.org/project/tensor-network-editor/} \\
\shortstack[l]{\texttt{quantum-circuit-}\\\texttt{drawer}} & 1.1.1 & 18 May 2026 & \url{https://github.com/DOKOS-TAYOS/quantum-circuit-drawer} & \url{https://pypi.org/project/quantum-circuit-drawer/} \\
\bottomrule
\end{tabular}
\endgroup
\label{tab:availability}
\end{table}
\footnotetext[3]{PyPI sdist SHA256 digests:\\
\texttt{tensor-network-visualization}: \pkgsha{05040c2754121c51708a1a81bb8649aec68b04e65cc119dde572aeb64efebc4f}\\
\texttt{tensor-network-editor}: \pkgsha{900b6b6c5b7285fb00742d772e02a122173f36c94ad566deea016f844b34ae03}\\
\texttt{quantum-circuit-drawer}: \pkgsha{ba8d8f16a10a87637d7db00b27d3a161c9e7626aff6e487abd42aacb5413a50d}.}

\section{AI Usage Disclosure}

Generative AI systems, including OpenAI ChatGPT models, were used as drafting and editorial assistants during the preparation of this manuscript.
The author defined the paper's ideas, scope, structure, technical framing, package knowledge, and the source material derived from the libraries themselves, including the conceptual content and the screenshots used to document package behavior.
The AI assistance was used primarily to draft prose, prepare tables and diagrams, suggest text variants, and support iterative reviews for missing explanations or weak transitions.
All substantive prompts, corrections, factual steering, error correction, redrafting instructions, and final acceptance decisions were provided by the author, who reviewed and revised the generated material throughout.
The author also used AI-assisted review to identify missing explanations, weak transitions, and possible inconsistencies.
The author takes responsibility for the manuscript's claims, interpretation, and final wording.

\section*{Acknowledgements}

This work was developed within the When Physics initiative ``When Physics Becomes Science'', which supports open scientific tools and educational material for quantum and mathematical physics.

\bibliographystyle{unsrt}
\bibliography{references}

\end{document}